\documentclass[aps,preprint]{revtex4}%
\usepackage{amsfonts}
\usepackage{amsmath}
\usepackage{amssymb}
\usepackage{graphicx}%
\begin{document}
\preprint{CTP-SCU/2020003}
\title{Validity of Thermodynamic Laws and Weak Cosmic Censorship for AdS Black Holes
and Black Holes in a Cavity}
\author{Peng Wang$^{a}$}
\email{pengw@scu.edu.cn}
\author{Houwen Wu$^{a}$}
\email{iverwu@scu.edu.cn}
\author{Shuxuan Ying$^{a,b}$}
\email{ysxuan@cqu.edu.cn}
\affiliation{$^{a}$Center for Theoretical Physics, College of Physics, Sichuan University,
Chengdu, 610064, PR China}
\affiliation{$^{b}$Department of Physics, Chongqing University, Chongqing, 401331, China}

\begin{abstract}
By throwing a test charged particle into a Reissner-Nordstrom (RN) black hole,
we test the validity of the first and second laws of thermodynamics and weak
cosmic censorship conjecture (WCCC) with two types of boundary conditions,
i.e., the asymptotically anti-de Sitter (AdS) space and a Dirichlet cavity
wall placed in the asymptotically flat space. For the RN-AdS black hole, the
second law of thermodynamics is satisfied, and the WCCC is violated for both
extremal and near-extremal black holes. For the RN black hole in a cavity, the
entropy can either increase or decrease depending on the change in the charge,
and WCCC is satisfied/violated for the extremal/near-extremal black hole. Our
results indicate that there may be a connection between the black hole
thermodynamics and the boundary condition imposed on the black hole.

\end{abstract}
\maketitle
\tableofcontents

\section{Introduction}

Studying thermodynamic properties of black holes can have a deep impact upon
the understanding of quantum gravity. Penrose first noticed that a particle
can extract energy from a black hole with an ergosphere
\cite{IN-Penrose:1969pc}, which led to the discovery of the irreducible mass
\cite{IN-Christodoulou:1970wf,IN-Bardeen:1970zz,IN-Christodoulou:1972kt}. The
square of the irreducible mass of a black hole can be interpreted as the black
hole entropy \cite{IN-Bekenstein:1972tm,IN-Bekenstein:1973ur}. Later,
analogous to the laws of thermodynamics, the four laws of black hole mechanics
were proposed \cite{IN-Bardeen:1973gs}. With the advent of the AdS/CFT
correspondence \cite{IN-Maldacena:1997re}, there has been much interest in
studying thermodynamics and phase structure of AdS black holes
\cite{IN-Chamblin:1999tk,IN-Chamblin:1999hg,IN-Caldarelli:1999xj,IN-Cai:2001dz,IN-Kubiznak:2012wp,IN-Wei:2012ui,IN-Gunasekaran:2012dq,IN-Cai:2013qga,IN-Xu:2014kwa,IN-Frassino:2014pha,IN-Dehghani:2014caa,IN-Hennigar:2015esa,IN-Wang:2018xdz,IN-Wu:2019xse}%
, where some intriguing phase behavior, e.g., reentrant phase transitions and
tricritical points, was found to be present. It is also worth pointing out
that black holes can become thermally stable in AdS space since the AdS
boundary acts as a reflecting wall.

Along with the development of black hole thermodynamics, the weak cosmic
censorship conjecture (WCCC) was proposed to hide singularities by event
horizons \cite{IN-Penrose:1969pc}. So if the WCCC is valid, the singularities
cannot be seen by the observers at the future null infinity. To test the
validity of the WCCC, Wald constructed a gedanken experiment to destroy an
extremal Kerr-Newman black hole by overcharging or overspinning the black hole
via throwing a test particle into it \cite{IN-Wald}. Nevertheless, the
extremal Kerr-Newman black hole was shown to be incapable of capturing
particles with sufficient charge or angular momentum to overcharge or overspin
the black hole. Later, near-extremal charged/rotating black holes were found
to be overcharged/overspun by absorbing a particle
\cite{IN-Hubeny:1998ga,IN-Jacobson:2009kt,IN-Saa:2011wq}, and hence the WCCC
is violated. However, subsequent researches showed that the WCCC might be
still valid if the backreaction and self-force effects were considered
\cite{IN-Hod:2008zza,IN-Barausse:2010ka,IN-Barausse:2011vx,IN-Zimmerman:2012zu,IN-Colleoni:2015afa,IN-Colleoni:2015ena}%
. Since there is still a lack of the general proof of the WCCC, its validity
has been tested in various black holes
\cite{IN-Matsas:2007bj,IN-Richartz:2008xm,IN-Isoyama:2011ea,IN-Gao:2012ca,IN-Hod:2013vj,IN-Duztas:2013wua,IN-Siahaan:2015ljs,IN-Natario:2016bay,IN-Duztas:2016xfg,IN-Revelar:2017sem,IN-Sorce:2017dst,IN-Husain:2017cmj,IN-Crisford:2017gsb,IN-Gwak:2017kkt,IN-An:2017phb,IN-Ge:2017vun,IN-Yu:2018eqq,IN-Gwak:2018akg,IN-Gim:2018axz,IN-Chen:2018yah,IN-Zeng:2019jta,IN-Chen:2019nsr,IN-Gwak:2019asi,IN-Zeng:2019jrh,IN-Chen:2019pdj,IN-Wang:2019jzz,IN-Zeng:2019hux,IN-Hong:2019yiz,IN-Hu:2019rpw,IN-Hu:2019qzw,IN-Mu:2019bim,IN-He:2019fti,IN-Gan:2019jac,IN-Zeng:2019baw,IN-He:2019kws,IN-Hu:2019zxr,IN-Gwak:2019rcz,IN-Gan:2019ibg,IN-Hong:2020zcf,IN-Gwak:2020zht}%
. In particular, the thermodynamics and WCCC have been considered for a
Reissner-Nordstrom (RN)-AdS black via the charged particle absorption in the
normal and extended phase spaces \cite{IN-Gwak:2017kkt,IN-Zhang:2013tba}. In
the normal phase space, in which the cosmological constant is fixed, it showed
that the first and second laws of thermodynamics are satisfied while the WCCC
is violated even for an extremal RN-AdS black hole.

Instead of the AdS boundary, York showed that placing Schwarzschild black
holes inside a cavity, on the wall of which the metric is fixed, can make them
thermally stable \cite{IN-York:1986it}. Thermodynamics and phase structure of
RN black holes in a cavity were studied in a grand canonical ensemble
\cite{IN-Braden:1990hw} and a canonical ensemble
\cite{IN-Carlip:2003ne,IN-Lundgren:2006kt}. It was found that the
Schwarzschild and RN black holes in a cavity have quite similar phase
structure and transition to these of the AdS counterparts. Afterwards, various
black brane systems
\cite{IN-Lu:2010xt,IN-Wu:2011yu,IN-Lu:2012rm,IN-Lu:2013nt,IN-Zhou:2015yxa,IN-Xiao:2015bha}%
, a Gauss-Bonnet black hole \cite{IN-Wang:2019urm}, hairy black holes
\cite{IN-Sanchis-Gual:2015lje,IN-Sanchis-Gual:2016tcm,IN-Basu:2016srp,IN-Peng:2017gss}
and Boson stars
\cite{IN-Peng:2017squ,IN-Peng:2018abh,IN-Peng:2019qrl,IN-Peng:2020zmu} in a
cavity were extensively investigated, and it also showed that the behavior of
the gravity systems in a cavity is strikingly similar to that of the
counterparts in AdS gravity. However, we have recently studied phase structure
of Born-Infeld black holes enclosed in a cavity
\cite{IN-Wang:2019kxp,IN-Liang:2019dni} and thermodynamic geometry of RN black
holes in a cavity \cite{IN-Wang:2019cax}, and found their behavior has
dissimilarities from that of the corresponding AdS black holes. Note that
thermodynamics and critical behavior of de Sitter black holes in a cavity were
investigated in \cite{IN-Simovic:2018tdy,IN-Haroon:2020vpr}.

Although there have been a lot of work in progress on the thermodynamic laws
and WCCC for various black holes of different theories of gravity in
spacetimes with differing asymptotics, little is known about the second law of
thermodynamics and WCCC for a black hole enclosed in a cavity. Since RN-AdS
black holes and RN black holes in a cavity are thermally stable, they provide
appropriate scenarios to explore whether or not the thermodynamic laws and
WCCC are sensitive to the boundary condition of black holes. To this end, we
study the thermodynamic laws and WCCC for a RN black hole in a cavity in this paper.

The rest of this paper is organized as follows. In section \ref{Sec:RNAdSBH},
we review the discussion of the thermodynamic laws and WCCC for a RN-AdS black
hole to be self-contained and introduce the method used in this paper. The
thermodynamic laws and WCCC of a RN black hole in a cavity are then tested via
the absorption of a charged particle in section \ref{Sec:RNBHC}. We summarize
our results with a brief discussion in section \ref{Sec:Con}. For simplicity,
we set $G=\hbar=c=k_{B}=1$ in this paper.

\section{RN-AdS Black Hole}

\label{Sec:RNAdSBH}

In this section, we discuss the first and second laws of thermodynamics and
WCCC for a RN-AdS black hole by throwing a test particle into the black hole.
First, we consider the motion of a test particle of energy $E$, charge $q$ and
mass $m$ in a $4$-dimensional charged static black hole with the line
element,
\begin{equation}
ds^{2}=-f\left(  r\right)  dt^{2}+\frac{1}{f\left(  r\right)  }dr^{2}%
+r^{2}\left(  d\theta^{2}+\sin^{2}\theta d\phi^{2}\right)  ,
\end{equation}
and the electromagnetic potential $A_{\mu}$,
\begin{equation}
A_{\mu}=A_{t}\left(  r\right)  \delta_{\mu t}.
\end{equation}
We also suppose that the outermost horizon of the black hole is at $r=r_{+}$,
where $f\left(  r_{+}\right)  =0$. In \cite{IN-Wang:2019jzz}, the
Hamilton-Jacobi equation of the test particle was given by%
\begin{equation}
-\frac{\left[  E+qA_{t}\left(  r\right)  \right]  ^{2}}{f\left(  r\right)
}+\frac{\left[  P^{r}\left(  r\right)  \right]  ^{2}}{f\left(  r\right)
}+\frac{L^{2}}{r^{2}}=m^{2}\text{,} \label{eq:HJE}%
\end{equation}
where $L$ and $P^{r}\left(  r\right)  $ are the particle's angular momentum
and radial momentum, respectively. It was shown in
\cite{HJE-Benrong:2014woa,HJE-Tao:2017mpe} that $P^{r}\left(  r_{+}\right)  $
is finite and proportional to the Hawking temperature of the black hole. Eqn.
$\left(  \ref{eq:HJE}\right)  $ gives
\begin{equation}
E=-qA_{t}\left(  r\right)  +\sqrt{f\left(  r\right)  \left(  m^{2}+\frac
{L^{2}}{r^{2}}\right)  +\left[  P^{r}\left(  r\right)  \right]  ^{2}},
\label{eq:EqA}%
\end{equation}
where we choose the positive sign in front of the square root since the energy
of the particle is required to be a positive value
\cite{IN-Christodoulou:1970wf,IN-Christodoulou:1972kt}. At the horizon
$r=r_{+}$, the above equation reduces to%
\begin{equation}
E=q\Phi+\left\vert P^{r}\left(  r_{+}\right)  \right\vert , \label{eq:EP}%
\end{equation}
where $\Phi\equiv-A_{t}\left(  r_{+}\right)  $ is the electric potential of
the black hole. Eqn. $\left(  \ref{eq:EP}\right)  $ relates the energy of the
particle to its radial momentum and potential energy just before the particle
enters the horizon.

To check whether a particle can reach or exist near the black hole
horizon, we can rewrite eqn. $\left(  \ref{eq:EqA}\right)  $ as the radial
equation of motion,%
\begin{equation}
\left(  \frac{dr}{d\tau}\right)  ^{2}=\frac{\left[  E+qA_{t}\left(  r\right)
\right]  ^{2}}{m^{2}}-f\left(  r\right)  \left(  1+\frac{L^{2}}{m^{2}r^{2}%
}\right)  ,\label{eq:drdtau}%
\end{equation}
where we use $P^{r}\left(  r\right)  =mdr/d\tau$, and $\tau$ is the affine
parameter along the worldline. Note that a particle can exist in the region
where $\left(  dr/d\tau\right)  ^{2}\geq0$, and $dr/d\tau=0$ corresponds to a
turning point. Specifically for a particle existing at the event horizon, eqn.
$\left(  \ref{eq:drdtau}\right)  $ gives that $\left(  dr/d\tau\right)
^{2}|_{r=r_{+}}\geq0$, and hence $E\geq q\Phi$. Furthermore, if the particle
falls into the black hole, one has $\left(  dr/d\tau\right)  ^{2}>0$ at the
event horizon, which leads to $E>q\Phi$. In short, for a particle of energy
$E$ and charge $q$ around a black hole of potential $\Phi$, $E>q\Phi$ provides
a lower bound $E_{\text{low}}$ on $E$, which ensures that the particle exists
near the event horizon and gets absorbed by the black hole.

For a RN-AdS black hole, the metric function $f\left(  r\right)  $ and
electric potential $\Phi$ are%
\begin{equation}
f\left(  r;M,Q\right)  =1-\frac{2M}{r}+\frac{Q^{2}}{r^{2}}+\frac{r^{2}}{l^{2}%
}\text{ and }\Phi=\frac{Q}{r_{+}}, \label{eq:AdSmetric}%
\end{equation}
respectively, where $M$ and $Q$ are the mass and charge of the black hole,
respectively, and $l$ is the AdS radius. Here the parameters $M$ and $Q$ are
put explicitly as arguments of the function $f\left(  r\right)  $ for later
convenience. With fixed charge $Q$, the mass $M_{e}\left(  Q\right)  $ and
horizon radius $r_{e}\left(  Q\right)  $ of the extremal RN-AdS black hole are
determined by $f\left(  r_{e}\left(  Q\right)  ;M,Q\right)  =df\left(
r;M,Q\right)  /dr|_{r=r_{e}\left(  Q\right)  }=0$, which gives%
\begin{align}
M_{e}\left(  Q\right)   &  =\frac{\sqrt{6}l}{18}\left(  2+\sqrt{1+12Q^{2}%
/l^{2}}\right)  \sqrt{\sqrt{1+12Q^{2}/l^{2}}-1},\nonumber\\
r_{e}\left(  Q\right)   &  =\frac{l}{\sqrt{6}}\sqrt{\sqrt{1+12Q^{2}/l^{2}}-1}.
\end{align}
Like a RN black hole, when $M\geq$ $M_{e}\left(  Q\right)  $, the RN-AdS black
hole solution possesses the event horizon at $r=r_{+}\left(  M,Q\right)  $,
which is obtained by solving $f\left(  r;M,Q\right)  =0$
\cite{IN-Zhang:2013tba}. Otherwise, the event horizon disappears, and a naked
singularity emerges, which leads to the violation of the WCCC. If the event
horizon exists, one can define the black hole's temperature $T$ and entropy
$S$ as%
\begin{equation}
T=\frac{1}{4\pi}\frac{\partial f\left(  r;M,Q\right)  }{\partial r}%
|_{r=r_{+}\left(  M,Q\right)  }\text{ and }S=\pi r_{+}^{2}\left(  M,Q\right)
. \label{eq:AdSTandS}%
\end{equation}

Suppose one starts with an initial black hole of mass $M$ and charge $Q$ with
$M\geq$ $M_{e}\left(  Q\right)  $ and throws a test particle of energy $E\ll
M$ and charge $q\ll Q$ into the black hole. After the particle is absorbed,
the final configuration has mass $M+dM$ and charge $Q+dQ$. The energy and
charge conservation of the absorbing process gives%
\begin{equation}
dM=E\text{ and }dQ=q\text{,} \label{eq:AdSdMdQ}%
\end{equation}
where $E$ and $q$ are related via eqn. $\left(  \ref{eq:EP}\right)  $. If
$M+E\geq$ $M_{e}\left(  Q+q\right)  $, there exits an event horizon in the
final black hole solution, which hides the naked singularity. However for
$M+E<$ $M_{e}\left(  Q+q\right)  $, the naked singularity can be seen by
distant observers due to the absence of the event horizon.

We first check the first and second laws of thermodynamics for a RN-AdS black
hole during the absorption. In this case, the final black hole solution should
possess an event horizon at $r=r_{+}\left(  M,Q\right)  +dr_{+}$ in order that
thermodynamic variables are well defined.\ So the horizon radius, mass and
charge of the final black hole satisfy%
\begin{equation}
f\left(  r_{+}\left(  M,Q\right)  +dr_{+};M+dM,Q+dQ\right)  =0\text{,}%
\end{equation}
which leads to%
\begin{equation}
\frac{\partial f\left(  r;M,Q\right)  }{\partial r}|_{r=r_{+}\left(
M,Q\right)  }dr_{+}+\frac{\partial f\left(  r;M,Q\right)  }{\partial
M}|_{r=r_{+}\left(  M,Q\right)  }dM+\frac{\partial f\left(  r;M,Q\right)
}{\partial Q}|_{r=r_{+}\left(  M,Q\right)  }dQ=0\text{.}%
\end{equation}
Using eqns. $\left(  \ref{eq:EP}\right)  $, $\left(  \ref{eq:AdSmetric}%
\right)  $, $\left(  \ref{eq:AdSTandS}\right)  $ and $\left(  \ref{eq:AdSdMdQ}%
\right)  $, one finds that the above equation gives
\begin{equation}
\left\vert P^{r}\left(  r_{+}\left(  M,Q\right)  \right)  \right\vert
=TdS\text{.} \label{eq:AdSPT}%
\end{equation}
For an extremal black hole with $T=0$, since $P^{r}\left(  r_{+}\left(
M,Q\right)  \right)  \propto T$, eqn. $\left(  \ref{eq:AdSPT}\right)  $ is
trivial. Nevertheless, for a non-extremal RN-AdS black hole with $T>0$, the
variation of entropy is%
\begin{equation}
dS=\frac{\left\vert P^{r}\left(  r_{+}\left(  M,Q\right)  \right)  \right\vert
}{T}>0\text{,}%
\end{equation}
which means the second law of thermodynamics is satisfied. Moreover, plugging
eqn. $\left(  \ref{eq:AdSdMdQ}\right)  $ and $\left(  \ref{eq:AdSPT}\right)  $
into eqn. $\left(  \ref{eq:EP}\right)  $ yields the first law of
thermodynamics:%
\begin{equation}
dM=\Phi dQ+TdS\text{.}%
\end{equation}

To test the WCCC, we consider an extremal or near-extremal RN-AdS black hole
and check whether throwing a charged particle can overcharge the black hole.
To overcharge the black hole, the final configuration should exceeds
extremality:%
\begin{equation}
M+E<M_{e}\left(  Q+q\right)  ,
\end{equation}
which, together with eqn. $\left(  \ref{eq:EP}\right)  $, gives the
constraints on the energy of the particle%
\begin{equation}
E_{\text{low}}\equiv\frac{qQ}{r_{+}\left(  M,Q\right)  }<E<M_{e}\left(
Q+q\right)  -M\equiv E_{\text{up}}. \label{eq:AdSECon}%
\end{equation}
Since $q\ll Q$, we can expand $M_{e}\left(  Q+q\right)  $ and obtain
\begin{equation}
E_{\text{up}}\simeq M_{e}\left(  Q\right)  -M+M_{e}^{\prime}\left(  Q\right)
q+\frac{M_{e}^{\prime\prime}\left(  Q\right)  }{2}q^{2},
\end{equation}
where
\begin{equation}
M_{e}^{\prime}\left(  Q\right)  =\frac{Q}{r_{e}\left(  Q\right)  }\text{ and
}M_{e}^{\prime\prime}\left(  Q\right)  =\frac{\sqrt{\sqrt{1+12Q^{2}/l^{2}}-1}%
}{l\sqrt{2/3+8Q^{2}/l^{2}}}>0.
\end{equation}
If the initial black hole is extremal, the lower and upper bounds on $E$
become%
\begin{equation}
E_{\text{low}}=\frac{qQ}{r_{e}\left(  Q\right)  }\text{ and }E_{\text{up}%
}\simeq M_{e}^{\prime}\left(  Q\right)  q+\frac{M_{e}^{\prime\prime}\left(
Q\right)  }{2}q^{2}>E_{\text{low}}.
\end{equation}
So there always exists a test changed particle with $E_{\text{low}%
}<E<E_{\text{up}}$, which can overcharge the extremal RN-AdS black hole. For a
near-extremal RN-AdS black hole with $Q$ and $M=M_{e}\left(  Q\right)
+\epsilon^{2}$, the lower and upper bounds on $E$ become%
\begin{equation}
E_{\text{low}}\simeq M_{e}^{\prime}\left(  Q\right)  q-A\left(  Q\right)
q\epsilon\text{ and }E_{\text{up}}\simeq M_{e}^{\prime}\left(  Q\right)
q+\frac{M_{e}^{\prime\prime}\left(  Q\right)  }{2}q^{2}-\epsilon^{2},
\end{equation}
where $A\left(  Q\right)  >0$ is some function of $Q$, and $\epsilon$ is a
small parameter. To have $E_{\text{up}}>E_{\text{low}}$, we find
\begin{equation}
q\equiv a\epsilon>\frac{-A\left(  Q\right)  +\sqrt{A^{2}\left(  Q\right)
+2M_{e}^{\prime\prime}\left(  Q\right)  }}{M_{e}^{\prime\prime}\left(
Q\right)  }\epsilon. \label{eq:AdSNEQ}%
\end{equation}
The constraints $\left(  \ref{eq:AdSECon}\right)  $ gives the energy $E$ of
the particle,%
\begin{equation}
E=M_{e}^{\prime}\left(  Q\right)  a\epsilon+b\epsilon^{2}\text{ with
}-A\left(  Q\right)  a<b<\frac{M_{e}^{\prime\prime}\left(  Q\right)  }{2}%
a^{2}-1. \label{eq:AdSNEE}%
\end{equation}
Therefore, a charged particle with its charge and energy satisfying eqns.
$\left(  \ref{eq:AdSNEQ}\right)  $ and $\left(  \ref{eq:AdSNEE}\right)  $,
respectively, can overcharge the near-extremal black hole. In summary, WCCC is
always violated for extremal and near-extremal RN-AdS black holes.

\section{RN Black Hole in a Cavity}

\label{Sec:RNBHC}

In this section, we throw a charged particle into a RN black hole enclosed in
a cavity and test the first and second laws of thermodynamics and WCCC. We now
consider a thermodynamic system with a RN black hole enclosed in a cavity, the
wall of which is at $r=r_{B}$. The $4$-dimensional RN black hole solution is%
\begin{align}
ds^{2}  &  =-f\left(  r;M,Q\right)  dt^{2}+\frac{dr^{2}}{f\left(
r;M,Q\right)  }+r^{2}\left(  d\theta^{2}+\sin^{2}\theta d\phi^{2}\right)
\text{,}\nonumber\\
f\left(  r;M,Q\right)   &  =1-\frac{2M}{r}+\frac{Q^{2}}{r^{2}}\text{, }%
A=A_{t}\left(  r\right)  dt=-\frac{Q}{r}dt\text{,}%
\end{align}
where $M$ and $Q$ are the black hole charge and mass, respectively. The
Hawking temperature $T_{\text{BH}}$ of the black hole is given by%
\begin{equation}
T_{\text{BH}}=\frac{1}{4\pi}\frac{\partial f\left(  r;M,Q\right)  }{\partial
r}|_{r=r_{+}\left(  M,Q\right)  }=\frac{1}{4\pi r_{+}\left(  M,Q\right)
}\left(  1-\frac{Q^{2}}{r_{+}^{2}\left(  M,Q\right)  }\right)  ,
\label{eq:CavityTBH}%
\end{equation}
where $r_{+}\left(  M,Q\right)  =M+\sqrt{M^{2}-Q^{2}}$ is the radius of the
outer event horizon. Suppose that the wall of the cavity is maintained at a
temperature of $T$. It was shown in \cite{IN-Braden:1990hw} that the system
temperature $T$ can be related to the black hole temperature $T_{\text{BH}}$
as%
\begin{equation}
T=\frac{T_{\text{BH}}}{\sqrt{f\left(  r_{B};M,Q\right)  }}\text{,}
\label{eq:CavityT}%
\end{equation}
which means that $T,$ measured at $r=r_{B}$, is blueshifted from
$T_{\text{BH}}$, measured at $r=\infty$. The thermal energy $\mathcal{E}$ and
potential $\Phi$ of this system were \cite{IN-Braden:1990hw}%
\begin{align}
\mathcal{E}  &  =r_{B}\left[  1-\sqrt{f\left(  r_{B};M,Q\right)  }\right]
,\nonumber\\
\Phi &  =\frac{A_{t}\left(  r_{B}\right)  -A_{t}\left(  r_{+}\right)  }%
{\sqrt{f\left(  r_{B};M,Q\right)  }}. \label{eq:CavityEphi}%
\end{align}
The physical space of $r_{+}\left(  M,Q\right)  $ is bounded by%
\begin{equation}
r_{e}\left(  Q\right)  \leq r_{+}\left(  M,Q\right)  \leq r_{B}\text{,}
\label{eq:rBound}%
\end{equation}
where $r_{e}\left(  Q\right)  =Q$ is the horizon radius of the extremal black hole.

After we throw a particle of energy $E$ and charge $q$ into the RN black hole,
the thermal energy and charge of the system are changed from $\left(
\mathcal{E}\text{, }Q\right)  $ to $\left(  \mathcal{E+}d\mathcal{E}\text{,
}Q+dQ\right)  $. The energy and charge conservation give
\begin{align}
dQ  &  =q\text{,}\nonumber\\
d\mathcal{E}  &  =\frac{1}{\sqrt{f\left(  r_{B};M,Q\right)  }}\left(
dM-\frac{QdQ}{r_{B}}\right)  =E, \label{eq:CavitydQdE}%
\end{align}
where we use eqn. $\left(  \ref{eq:CavityEphi}\right)  $ to express
$d\mathcal{E}$ in terms of $dM$ and $dQ$. Here we assume that the radius
$r_{B}$ of the cavity is fixed during the absorption. Eqn. $\left(
\ref{eq:CavitydQdE}\right)  $ leads to the variation of the black hole mass
$M$,
\begin{equation}
dM=\sqrt{f\left(  r_{B};M,Q\right)  }E+\frac{qQ}{r_{B}}. \label{eq:CavitydM}%
\end{equation}
To discuss the thermodynamic laws, the final RN black hole after absorbing the
particle is assumed to possess an event horizon, which is located at
$r=r_{+}\left(  M,Q\right)  +dr_{+}$. Similar to the RN-AdS case, we have%
\begin{equation}
\frac{\partial f\left(  r;M,Q\right)  }{\partial r}|_{r=r_{+}\left(
M,Q\right)  }dr_{+}+\frac{\partial f\left(  r;M,Q\right)  }{\partial
M}|_{r=r_{+}\left(  M,Q\right)  }dM+\frac{\partial f\left(  r;M,Q\right)
}{\partial Q}|_{r=r_{+}\left(  M,Q\right)  }dQ=0\text{.}%
\end{equation}
Eqns. $\left(  \ref{eq:EP}\right)  $, $\left(  \ref{eq:CavityTBH}\right)  $,
$\left(  \ref{eq:CavityT}\right)  $ and $\left(  \ref{eq:CavitydM}\right)  $
then give
\begin{equation}
TdS=\left\vert P^{r}\left(  r_{+}\left(  M,Q\right)  \right)  \right\vert
+\frac{qQ}{r_{+}\left(  M,Q\right)  }-q\Phi, \label{eq:CavityTdS}%
\end{equation}
where $S=\pi r_{+}^{2}\left(  M,Q\right)  $ is the entropy of the system.
Using eqn. $\left(  \ref{eq:CavityEphi}\right)  $, we rewrite eqn. $\left(
\ref{eq:CavityTdS}\right)  $ as
\begin{equation}
TdS=\left\vert P^{r}\left(  r_{+}\left(  M,Q\right)  \right)  \right\vert
+\left(  \frac{1}{r_{+}}-\frac{1}{r_{+}\sqrt{f\left(  r_{B};M,Q\right)  }%
}+\frac{1}{r_{B}\sqrt{f\left(  r_{B};M,Q\right)  }}\right)  QdQ,
\label{eq:CavityTdSL}%
\end{equation}
where the prefactor of $QdQ$ is positive. It can show that, when $T=0$ (i.e.,
$Q=M$), both sides of $\left(  \ref{eq:CavityTdSL}\right)  $ are zero, which
cannot provide any information about $dS.$ For a non-extremal black hole, eqn.
$\left(  \ref{eq:CavityTdSL}\right)  $ gives that the entropy increases when
$dQ>0$. However when $dQ<0$, the entropy can increases or decrease depending
the value of $dQ$. So the second law of thermodynamics is indefinite for a RN
black hole in a cavity. Substituting eqn. $\left(  \ref{eq:EP}\right)  $ into
eqn. $\left(  \ref{eq:CavityTdS}\right)  $, we obtain%
\begin{equation}
d\mathcal{E}=\Phi dQ+TdS\text{,}%
\end{equation}
which is the first law of thermodynamics.

To overcharge a RN black hole of mass $M$ and charge $Q$ in a cavity by a test
particle of energy $E$ and charge $q$, the mass $M+dM$ and charge $Q+q$ of the
final configuration must satisfy%
\begin{equation}
M+dM<Q+q, \label{eq:CavityMQ}%
\end{equation}
which, due to eqn. $\left(  \ref{eq:CavitydM}\right)  $, puts an upper bound
on $E$,%
\begin{equation}
E<E_{\text{up}}\equiv\frac{Q+q-M-\frac{qQ}{r_{B}}}{\sqrt{f\left(
r_{B};M,Q\right)  }}.
\end{equation}
On the other hand, eqn. $\left(  \ref{eq:EP}\right)  $ also puts a lower bound
on $E$,%
\begin{equation}
E>\frac{qQ}{r_{+}\left(  M,Q\right)  }\equiv E_{\text{low}}\text{.}%
\end{equation}
For an extremal black hole with $M=Q$, we find%
\begin{equation}
E_{\text{up}}=E_{\text{low}}=q\text{,}%
\end{equation}
which indicates that the extremal black hole cannot be overcharged.
Considering a near-extremal RN black hole with $Q$ and $M=Q+\epsilon^{2}$, one
can show that%
\begin{equation}
E_{\text{low}}\simeq q-\frac{\sqrt{2}q\epsilon}{\sqrt{Q}}\text{ and
}E_{\text{up}}\simeq q-\frac{\epsilon^{2}}{1-Q/r_{B}}.
\end{equation}
If $E_{\text{up}}>E_{\text{low}}$, the charge $q$ of the test particle should
be bounded from below,%
\begin{equation}
q\equiv a\epsilon>\frac{\epsilon}{\left(  1-Q/r_{B}\right)  }\sqrt{\frac{Q}%
{2}}. \label{eq:Cavityq}%
\end{equation}
Moreover, the corresponding energy $E$ of the particle is
\begin{equation}
E=a\epsilon+b\epsilon^{2}\text{ with }-\frac{\sqrt{2}a}{\sqrt{Q}}<b<-\frac
{1}{1-Q/r_{B}}. \label{eq:CavityE}%
\end{equation}
So a test particle with $\left(  q,E\right)  $ in the parameter region
$\left(  \ref{eq:Cavityq}\right)  $ and $\left(  \ref{eq:CavityE}\right)  $
can overcharge the near-extremal RN black hole in a cavity, which invalidates
the WCCC.

\section{Conclusion}

\label{Sec:Con}

In this paper, via absorbing a test charged particle, we calculated the
variations of thermodynamic quantities of RN black holes with two boundary
conditions, namely the asymptotically AdS boundary and the Dirichlet boundary
in the asymptotically flat spacetime. With these variations, we checked the
first and second laws of thermodynamics and WCCC in these two cases. Our
results are summarized in Table \ref{tab:1}. In the limit of $l\rightarrow
\infty$, a RN-AdS black hole becomes a RN black hole. Similarly, a RN black
hole in a cavity also reduces to a RN black hole when $r_{B}\rightarrow\infty
$. We find that taking the limits of our results in both AdS and cavity cases
gives the same result about the validity of the thermodynamic laws and WCCC
for a RN black hole, which is also presented in Table \ref{tab:1}.

\begin{table}[tbh]
\centering%
\begin{tabular}
[c]{|p{0.6in}|p{1.6in}|p{1.95in}|p{1.95in}|}\hline
& RN-AdS BH & RN BH in a cavity & RN BH\\\hline
1st Law & Satisfied. & Satisfied. & Satisfied.\\\hline
2nd Law & Satisfied. & Satisfied for $dQ>0$. Indefinite for $dQ<0$. &
Satisfied.\\\hline
WCCC & Violated for extremal and near-extremal BHs. & Satisfied for extremal
BH. Violated for near-extremal BH. & Satisfied for extremal BH. Violated for
near-extremal BH.\\\hline
\end{tabular}
$\ \ $\caption{{\small Results for the first and second laws of thermodynamics
and weak cosmic censorship conjectures (WCCC), which are tested for a RN-AdS
black hole, a RN black hole in a cavity and a RN black hole by absorbing a
test charged particle.}}%
\label{tab:1}%
\end{table}

In \cite{IN-Carlip:2003ne,IN-Lundgren:2006kt}, thermodynamic and phase
structure of a RN-AdS black hole and a RN black hole in a cavity were shown to
be strikingly similar. However, it was found that thermodynamic geometry in
these two cases behaves rather differently \cite{IN-Wang:2019cax}, which
implies that there may be a connection between the black hole microstates and
the boundary condition. In this paper, we showed that the validity of the
second law of thermodynamic and WCCC in the AdS and cavity cases are quite
different, which further motivates us to explore the connection between the
internal microstructure of black holes and the boundary condition.

To understand the scale of energy required for a particle to overcharge a
black hole, we can convert physical quantities in Plack units to SI units. In
fact, for a particle of charge $q$ and energy $E$ in Plack units, the charge
and the energy in SI units are $qq_{p}$ and $Em_{p}c^{2}$, respectively, where
$q_{p}\equiv\sqrt{4\pi\epsilon_{0}\hbar c}=e/\sqrt{\alpha}$ is the Planck
charge, $\epsilon_{0}$ is the permittivity of free space, $e$ is the
elementary charge, $\alpha$ is the fine structure constant, and $m_{p}%
=\sqrt{\hbar c/G}$ is the Planck mass. From eqns. $\left(  \ref{eq:AdSNEQ}%
\right)  $, $\left(  \ref{eq:AdSNEE}\right)  $, $\left(  \ref{eq:Cavityq}%
\right)  $ and $\left(  \ref{eq:CavityE}\right)  $, we find that a test
particle that can overcharge a charged black hole should have
\begin{equation}
E\sim q\text{ (in Planck units),}%
\end{equation}
which leads to%
\begin{equation}
E\sim\frac{\left\vert q\right\vert }{e}\left(  \sqrt{\alpha}m_{p}c^{2}\right)
\sim\frac{\left\vert q\right\vert }{e}\times10^{18}%
%TCIMACRO{\unit{GeV}}%
%BeginExpansion
\operatorname{GeV}%
%EndExpansion
\text{ (in SI units).}%
\end{equation}
For example, if one throws an ionized Hydrogen nuclei of $m\sim1%
%TCIMACRO{\unit{GeV}}%
%BeginExpansion
\operatorname{GeV}%
%EndExpansion
/c^{2}$ and $q=e$ to overcharge a charged black hole, the nuclei is
ultrarelativistic with an enormous kinetic energy $\sim10^{18}%
%TCIMACRO{\unit{GeV}}%
%BeginExpansion
\operatorname{GeV}%
%EndExpansion
$ to overcome the electrostatic repulsion between the particle and the black hole.

In this paper, we discussed the WCCC in the test limit, in that the
interaction between test particles and the black hole background is neglected.
Although this method is simple and straightforward, it is subject to several
limitations. For example, our results showed that the final mass of the black
hole needs to have a second-order correction in $q$ to overcharge the black
hole. However, if one calculates the mass consistently to order $q^{2}$, the
test limit, which is valid only to linear order in $q$, is not enough, and
hence all second-order effects, e.g., self-force and finite size effects,
should be considered. In \cite{IN-Sorce:2017dst}, a general formula for the
full second-order correction to mass was proposed, and it was found that the
WCCC is valid for Kerr-Newman black holes up to the second-order perturbation
of the matter fields. In this new version of the gedanken experiment, the WCCC
was tested and found to be valid for various black holes
\cite{IN-An:2017phb,IN-Ge:2017vun,Con-Jiang:2019ige,Con-Jiang:2019vww}. In
particular, the WCCC was investigated for RN-AdS black holes in the extended
phase space in \cite{Con-Wang:2019bml}, which showed that the WCCC cannot be
violated for RN-AdS black holes under the second-order approximation of the
matter field perturbations. On the other hand, we used the test limit to study
the WCCC for RN-AdS black holes in the normal phase space with fixed
cosmological constant, and found that the WCCC is violated. Apart from
different phase spaces considered, differences between our results and these
in \cite{Con-Wang:2019bml} suggest that corrections beyond the test limit can
play an important role in the analysis of the WCCC for charged AdS black holes
and black holes in a cavity. We leave this for future work.

\bigskip

\begin{acknowledgments}
We are grateful to Haitang Yang and Zhipeng Zhang for useful discussions and
valuable comments. This work is supported in part by NSFC (Grant No. 11875196,
11375121 and 11005016).
\end{acknowledgments}

\end{document}